\documentclass[letterpaper,11pt]{article}
\pdfoutput=1 

\usepackage{jheppub} 
\usepackage[utf8]{inputenc}
\usepackage[T1]{fontenc} 
\usepackage[dvipsnames]{xcolor}
\usepackage{multirow}
\usepackage{slashed}
\usepackage[normalem]{ulem}
\usepackage{sidecap}
\usepackage{mathtools}
\usepackage{caption}
\usepackage{subcaption}
\usepackage[toc]{appendix}
\usepackage{enumerate}

\title{Probing the flavor-specific scalar mediator for the muon $(g-2)$ deviation, the proton radius puzzle and the light dark matter production}
\date{\today}
\author[1,2]{Bin Zhu,}
\author[1]{Xuewen Liu}

\affiliation[1]{Department of Physics, Yantai University, Yantai 264005, China}
\affiliation[2]{Department of Physics, Chung-Ang University, Seoul 06974, Korea}

\emailAdd{zhubin@mail.nankai.edu.cn}
\emailAdd{xuewenliu@ytu.edu.cn}

\abstract{
Flavor-specific scalar bosons exist in various Standard Model extensions and couple to a single generation of fermions via a global flavor symmetry breaking mechanism. Given this strategy, we propose a MeV flavor-specific scalar model in dimension-$5$ operator series, which explains the muon g-2 anomaly and proton radius puzzle by coupling with the muon and down-quark at the same time. The framework is consistent with the null result of high-intensity searches. Specifically, the supernova constraints for muon couplings become weakened by including the contribution of down-quark interaction. The parameter space for explaining muon $g-2$ discrepancy is available when $10\%$ energy deposition is required in the energy explosion process in the supernova, but this is ruled out by the $1\%$ energy deposition requirement. We also investigate the searches for mediator and dark matter and the resulting constraints on viable parameter space such as nuclear physics constraints, direct detection for light boosted dark matter, and possible CMB constraints. When compared to conventional dark matter production, light dark matter production has two additional modifications: bound state formation and early kinetic equilibrium decoupling. We are now looking into the implications of these effects on the relic density of light dark matter.
}

\begin{document}
\maketitle

\section{Introduction}

Muon g-2 is an extremely sensitive probe for new physics models.
Fermilab recently released the first results of the muon g-2 experiment, which confirms the Brookhaven results on the long-standing anomaly of the anomalous magnetic moment of the muon, namely the discrepancy between the experimental and theoretical values of the anomalous magnetic moment ~\cite{Bennett:2006fi, Tanabashi:2018oca, Davier:2017zfy, Blum:2018mom, Keshavarzi:2018mgv, Davier:2019can, Aoyama:2020ynm}. The new result further strengthens the significance to be about  $4.2~\sigma$ deviation~\cite{PhysRevLett.126.141801},
\begin{equation}
\label{muon} \Delta a_\mu = a_\mu^{\rm exp}-a_\mu^{\rm th} = (2.51\pm 0.59)\times 10^{-9}~.
\end{equation}

This anomaly could be explained by a muonic force mediated by a scalar or vector boson in the MeV range. Furthermore, laser spectroscopy of muonic hydrogen~\cite{Pohl:2010zza} yielded a proton charge radius that does not agree, at about $5\sigma$ level, with the value obtained from electron-proton scattering and electron hydrogen spectroscopy~\cite{Mohr:2008fa, Mohr:2015ccw}. This discrepancy is the proton radius puzzle. It is yet another basic motivation for muonic force. Although some recent data on $ep$ scattering~\cite{Xiong:2019umf} and H spectroscopy~\cite{Beyer:2017gug, Bezginov:2019mdi} remove discrepancy with muonic hydrogen results to some extent, a few recent experimental collaborations ~\cite{Fleurbaey:2018fih, Mihovilovic:2019jiz} have reported $r_p$ values that persistently agree with the CODATA $2014$ $r_p$ value. As a result, the proton charge radius puzzle has been ameliorated but not resolved. We note that, 
\begin{equation}
    r_p^{(e)}=0.8751 \pm 0.0061~ {\rm fm}, \quad r_p^{(\mu)}=0.84087 \pm 0.00039~ {\rm fm}.
\end{equation}

The existence of new force~\cite{Essig:2013lka,Padley:2015uma,Alexander:2016aln,Battaglieri:2017aum,Das:2017ski,Sabatta:2019nfg,DAgnolo:2020mpt,Su:2020zny} mediated by a low-mass particle with very weak coupling to SM particles is the main target at the intensity frontiers~\cite{Alekhin:2015byh} as well as nuclear physics The resulting parameter bounds are shown in the context of a flavor-specific scalar model~\cite{Batell:2017kty, Batell:2018fqo, Krnjaic:2019rsv}. We suggest that flavor-specific scalar $\phi$ may be the relaxation process that resolves both discrepancies at the MeV scale. In general, the scalar mediator model is characterized by the Higgs portal interaction ~\cite{Krnjaic:2015mbs}. However, we adapt the effective field theory approach, $d=5$ operators, in which the mediator $\phi$ couples to both the muon and the down-quark at the same time. Higher-dimensional operators are suppressed by UV cut-off $M$. The spurion analysis enables a mediator to break flavor university and couple to a single generation of fermions. As a result, using this model-independent approach, we can investigate the multi-scale problems of relating high precision measurements with muons.

The two-photon couplings induced by the triangle loops involving both muon and down-quark result in the most sensitive constraint for the MeV scale scalar. In particular, the effective two-photon coupling $\phi F^{\mu\nu}F_{\mu\nu}$~\cite{Dobrich:2015jyk, Dolan:2017osp} in the supernova leads to the most restrictive constraint via diffuse gamma-ray. To remove the diffuse gamma-ray bound, we expect the scalar to decay within the radius of the progenitor star, resulting in an energy explosion bound instead. In this paper, we demonstrate that the muon $g-2$ discrepancy remains plausible if we require $10\%$ energy deposition in the explosion process.
We also note that the conventional energy loss bound from the supernova is insignificant for the muon $g-2$ discrepancy. Finally, when the mediator couples to a down-quark rather than an up-quark, the associated rare Kaon decay constraints in E949~\cite{Artamonov:2009sz} vanish.

From a perspective of dark matter, the calculations on $\Delta N_{\mathrm{eff}}$ show that dark matter is heavier than MeV. We will discuss the implication of light dark matter relic density produced by the $\chi\chi\rightarrow \phi\phi$ process together with Sommerfeld correction. Although the annihilation process is P-wave, the corresponding bound state formation cross-section is S-wave, introducing CMB constraints. Furthermore, direct detection loses sensitivity to light-dark matter because it does not produce efficient recoil energy in the noble liquid detector. Some astrophysical sources can boost light dark matter into a fast-moving component that recovers detector sensitivity. For example, cosmic-ray can scatter with dark matter in the halo~\cite{Bringmann:2018cvk, Ema:2018bih}. The resulting dark matter has enough kinetic energy to pass through the detector's threshold. As a result, this mechanism can limit our parameter space for the dark sector.

This paper is structured as follows: we begin in section \ref{sec:model} by presenting the flavor-specific scalar model. In section \ref{sec:muon-proton}, We explain both the Muon $(g-2)$ and proton charge radius anomalies. Furthermore, we examine the parameter space constraints derived from nuclear physics and various experimental data. In section \ref{sec:DM}, we investigate the DM production. As a proof of principle of the mass hierarchy $m_{\phi}<m_{\chi}$ from $\Delta N_{\mathrm{eff}}$ constraint, We will then use it to compute the relic density of dark matter. The relic density calculation belongs to the secluded dark matter scenario because the cross-section of direct annihilation into muons and photons is smaller than that into mediators $\chi\chi\rightarrow \phi\phi$~\cite{DAgnolo:2020mpt}. However, there are two possible modifications of the secluded dark matter. The underlying assumption on kinetic equilibrium relies upon the elastic scattering process between dark matter and mediator. If it is insufficient, early kinetic decoupling may occur, affecting the relic density. Another possible modification comes from the light mediator, which allows the bound state to exist in the cosmological evolution of dark matter. Its presence also has an impact on relic density. We will investigate it in detail in section~\ref{sec:DM}. Finally, the last section is devoted to our conclusion.

\section{Muon-philic and down-philic scalar mediator}
\label{sec:model}

Identifying the dark sector has become mainstream in physics society. There is widespread agreement that dark matter is a sub-GeV particle with a mediator that connects dark matter and Standard Model particles. Despite the fact that which types of mediators hypothesis has yet to be confirmed, we prefer the mediator a real scalar because other types of mediators cannot account for the proton radius puzzle \cite{Liu:2016qwd}. Therefore, we describe the dark sector-SM interaction through the model-independent dimension-$5$ operators

\begin{equation}
-\mathcal{L}_{\mathrm{d}=5}=\frac{C_{u}^{ij}}{M} \phi H^{c} \bar{Q}_{L}^{i} u_{R}^{j}+\frac{C_{d}^{ij}}{M} \phi H \bar{Q}_{L}^{i} d_{R}^{j}+\frac{C_{e}^{ij}}{M} \phi H E_{L}^{i} e_{R}^{j}+ h.c.
\label{eqn:operator}
\end{equation}

Indeed there are other dimension-5 operators i.e. $\partial_{\mu}\phi \bar{q}\gamma^{\mu}q$. After a field redefinition, they are equivalent to equation~\ref{eqn:operator}. In addition, the mediator $\phi$ could couple to neutrino via $C_{\nu}/M^2 \phi (HL)^2$. It is a dimension-6 operator, so it is insignificant 
unless an unnatural hierarchy of Wilson coefficients holds such as $C_{\nu}\sim 10^{10}C_{q}$.

The implications of such operators in equation~\ref{eqn:operator} for low energy phenomenologies such as the muon $(g-2)$ and the proton radius puzzle will be discussed. Here $i$ and $j$ stand for generation indices of quarks and leptons. Thus the Wilson coefficients $C_{u,d,e}$ are generic $3\times 3$ matrix. Some assumptions on global symmetry breaking of flavor structure are imposed from the standpoint of flavor violation constraint. For example, a breaking  $U(3)_Q\times U(3)_D\rightarrow U(1)_d\times U(2)_{sbL}\times U(2)_{sbR}$ can assure only $C_{d}^{11}$ is non-vanishing for quark Wilson coefficient $C_{u,d}$. Employing symmetry breaking pattern for leptons can yield a residue $U(1)_{\mu}$. Only $C_e^{22}$ is non-vanishing in this sense. We should point out that our analysis is merely a hypothesis that relies on UV completion.
For example, the simplest UV completion is to take SUSY breaking spurion-sgoldsitino~\cite{Liu:2020ser} as force carrier or other SUSY contribution~\cite{Endo:2021zal,Cao:2021lmj,Abdughani:2021pdc}.

After the electroweak symmetry breaking, the relevant portal interaction with SM from dimension-5 operator~\ref{eqn:operator} becomes
\begin{equation}
   - \mathcal{L}_{\mathrm{Portal}}=g_d \phi \bar{d}_L d_R+ g_{\mu}\phi \bar{\mu}_L \mu_R+h.c.,
\end{equation}
where $ g_d=C_d^{11}v/M,\quad g_{\mu}=C_e^{22}v/M$. In contrast to other approaches to the MeV force scenario, the novelty of our model is that proton and neutron couplings are uniquely fixed by $g_d$:
\begin{equation}
    g_p=0.039 g_d \frac{m_p}{m_d},\quad g_n=0.049 g_d\frac{m_n}{m_d}.
\end{equation}
For simplicity, we consider the dark matter to be a Dirac fermion and couples to the mediator $\phi$ in the following interaction,  
\begin{equation}
\mathcal{L}_{\mathrm{Dark}}=i \bar{\chi} \gamma^{\mu} \partial_{\mu} \chi-m_{\chi} \bar{\chi} \chi-g_{\chi} \bar{\chi} \chi \phi. 
\label{eqn:Dark}
\end{equation}

In the absence of other interactions and assuming the dark matter is heavier than the mediator, for $m_{\chi}>m_\phi > 2m_\mu$,
the dominant decay is $\phi \to \mu^+\mu^-$ with partial width

\begin{equation}
\Gamma_{\phi \to \mu^+\mu^-} = \frac{g_{\mu}^2 m_\phi}{8\pi} \left( 1 - \frac{4m_\mu^2}{m_\phi^2}\right)^{3/2}. 
\end{equation}

The beam dump experiment is also sensitive to decay channel into pions when $m_{\phi}>2m_{\pi}$. Because the proton radius puzzle favors a MeV-scale mediator, this channel does not exist. For $m_\phi < 2m_\mu$, the dominant channel is $\phi \to \gamma \gamma$ through a muon and down quark loop,

\begin{equation}
\Gamma_{\phi \rightarrow \gamma \gamma}=\frac{\alpha^{2}g_{\mu}^{2} m_{\phi}^{3}}{144 \pi^{3} m_{\mu}^{2}}\left|F_{1 / 2}\left(\frac{4 m_{\mu}^{2}}{m_{\phi}^{2}}\right)\right|^{2}+ \frac{\alpha^{2} N_{c}^{2} Q_{d}^{4} g_{d}^{2} m_{\phi}^{3}}{144 \pi^{3} m_{d}^{2}}\left|F_{1 / 2}\left(\frac{4 m_{d}^{2}}{m_{\phi}^{2}}\right)\right|^{2},
\end{equation}
where the loop function is

\begin{equation}
F_{1 / 2}(\tau)=\frac{3 \tau}{2}\left[1+(1-\tau)\left(\sin ^{-1} \frac{1}{\sqrt{\tau}}\right)^{2}\right].
\end{equation}

The visible decay final state $\phi\rightarrow \gamma\gamma,\mu\mu$ plays a crucial role in increasing the projection for the NA62~\cite{Lazzeroni:2012cx} search for $K\rightarrow \mu\nu\phi$. The supernova constraints exclude the majority of the available parameter space for muonic force. However, because our model has additional coupling with down-quarks, we can shift the excluded region to explain the muon $g-2$ discrepancy.

\section{The muon (g-2), proton charge radius and constraints}
\label{sec:muon-proton}

At one loop level, the scalar mediator contributes to the anomalous magnetic moment of the muon. The shift due to the vertex correction is~\cite{Jackiw:1972jz}

\begin{align}\label{eq:g-2}
\Delta a_\mu=\frac{g_\mu^2 }{8\pi^2}\int_0^1dz\frac{(1-z)^{2}(1+z)}{(1-z)^{2}+(m_\phi/m_\mu)^{2}z}.
\end{align}

\begin{figure}[!htbp]
\centering
\includegraphics[width=0.65\textwidth]{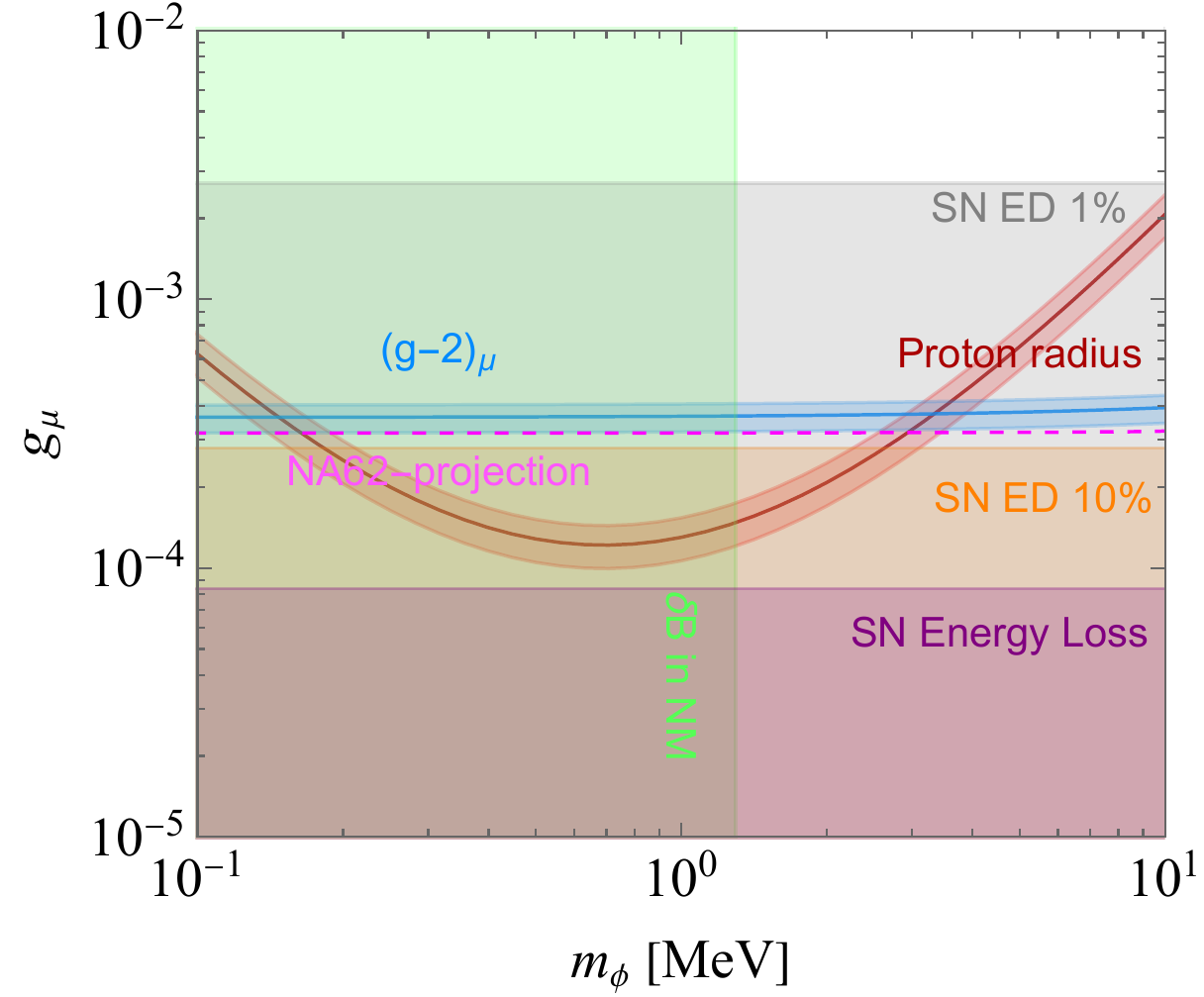}
\caption{
    Muon $(g-2)$ anomaly, proton radius puzzle and experimental constraints,
	plotted as functions of the scalar mass $m_\phi$ and coupling $g_\mu$.
	Here we set $g_d=2\times10^{-4}$.
	Blue: a region that can explain the $(g-2)_\mu$ anomaly in $2\sigma$ deviation.
	The blue line is the measured central value of the $\Delta a_\mu$ data.
	Red: region in which the proton radius puzzle can be solved.
	The solid line corresponds to the central value in equation~\ref{eq:deltaEvalue}, while the shaded regions include the 2$\sigma$ compatible values.
	Green: nuclear matter excluded regions due to nucleon binding energy. (indicated with ``$\delta B $ in NM''),
	Purple: by supernova energy loss (``SN Energy Loss''),
	Orange: by supernova explosion energy with 10\% deposition (``SN ED 10\%"),
	Gray: by supernova explosion energy with 1\% deposition (``SN ED 1\%'').
	%
	%
	The magenta dashed line corresponds to the NA62 projection for $K^+ \to \mu^+ + {\rm invisible}$ searching.
}
\label{fig:gmu}
\end{figure}

\begin{figure}[!htbp]
\centering
\includegraphics[width=0.65\textwidth]{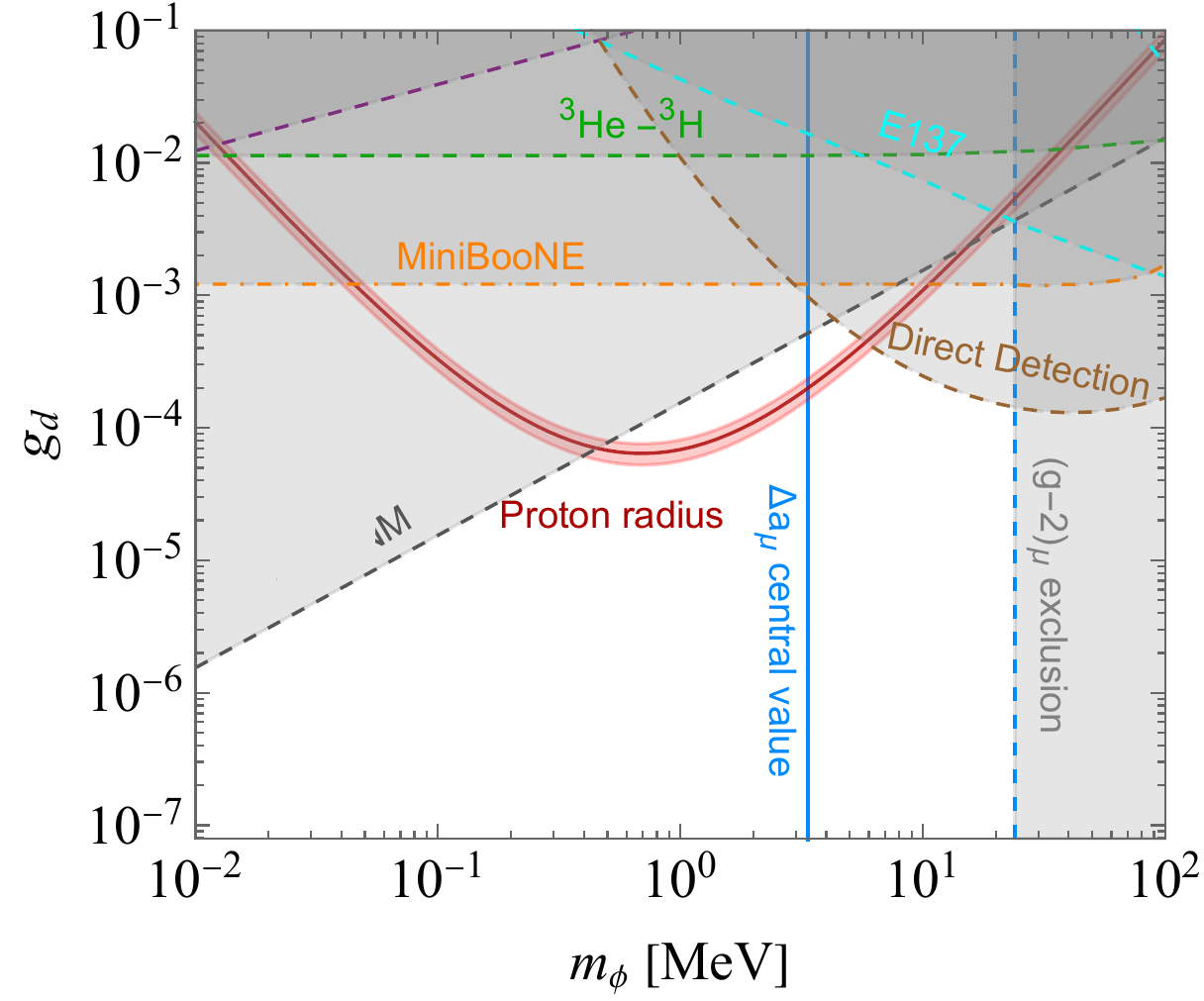}
\caption{
    Muon $(g-2)$ anomaly, proton radius anomoly, and experimental constraints 
    as functions of scalar mass $m_\phi$ and coupling $g_d$. 
	Here we set $g_\mu=3.8\times10^{-4}$.
	Blue lines:
	The solid blue line is the measured central value of the $\Delta a_\mu$ data.
	The dashed blue line is exclusion limits on scalar mass.
	Red: a region where can address the proton radius puzzle.
	The solid line represents the central value, while the shaded areas represent the  2$\sigma$ compatible values.
	Gray regions: excluded regions by various constraints.
	The dashed lines in black, green, purple, orange, cyan, and brown correspond to the constraints imposed by nucleon binding energy in the nuclear matter ($\delta B$ in NM).
	$^3$He- $^3$H binding energy difference ( $^3$He- $^3$H),
	neutron-$^{208}$Pb Scattering (N-N), MiniBooNE, E137, and the DM direct detection.
	}
\label{fig:gd}
\end{figure}

Using this new effect, we were able to close the gap between experimental data and theoretical prediction within $2$ standard deviations, as shown in blue in Figs.  \ref{fig:gmu} and \ref{fig:gd}. The Favored values of muonic coupling $g_\mu$ are around $10^{-4}$.

In Fig. \ref{fig:gd}, we set $g_\mu=3.8\times10^{-4}$ according to allowed parameter space in Fig.~\ref{fig:gmu}, which also meets the experimental constraints, as will be discussed further below. To obtain the central value of $\Delta a_\mu$, the mediator mass is solved to be 3.4 MeV, shown by the blue vertical line. Also, the range of $m_\phi>20$ MeV is out of the data error at the time.

Spectroscopy in Hydrogen-like atoms can
be changed by the finite-sized proton, which induces the Lamb to shift $\Delta E_{\rm Lamb}$ between the 2S and 2P energy levels. Then the measurements of the Lamb shift can deduce the proton charge radius.
The presence of the new scalar causes an extra contribution to the Lamb shift in muonic hydrogen, which could account for the deviation in proton size measurements. Typically, the Yukawa potential introduces a new attractive force that alters the $\mu-p$ system, 

\begin{equation}
V(r)=-\frac{1}{4\pi}g_{\mu} g_{p} e^{-m_\phi r}/r.
\end{equation}

As a result, the additional contribution to the Lamb shift $\Delta E_{\rm Lamb}$ in muonic hydrogen is given by \cite{Barger:2010aj,TuckerSmith:2010ra,Miller:2015yga, Perelstein:2020suc}

\begin{align}\label{eq:Lamb shift}
\delta E_{\rm Lamb}^{\ell\rm N} &=\left<2S|V(r)|2S\right>-\left<2P|V(r)|2P\right> \\
                       &=-\frac{g_\mu}{8\pi a_{\ell\rm N}}[Z g_p+(A-Z) g_n] f(a_{\ell\rm N}m_\phi),
                       \end{align}
where $f(x)=x^2/(1+x)^4$, with $a_{\ell\rm N}=(Z\alpha\,m_{\ell\rm N})^{-1}$ the Bohr radius and $m_{\ell\rm N}$ is the lepton-nucleus system's reduced mass. $A$ and $Z$ are the atom number and proton number in an atom, respectively.
In the following, we use
\begin{align}\label{eq:deltaEvalue}
\delta E_{\rm Lamb}^{\mu\rm H}=-0.307(56)~\rm meV
\end{align}
for fitting our model parameters within two standard deviations. Using this energy shift, we can calculate $g_\mu$ and $g_d$ as functions of $m_\phi$. The red-hued areas in Figs.~\ref{fig:gmu} and~\ref{fig:gd} depict the central value and error band of $\delta E_{\rm Lamb}^{\mu\rm H}$. To demonstrate the results explicitly, in Fig.~\ref{fig:gmu} we set $g_d=2\times10^{-4}$ which is allowed by various experimental constraints.
In Figs.~\ref{fig:gmu} and \ref{fig:gd}, We discover that there is enough room to accommodate both anomalies at the same time, and the mediator mass of $3.4$ MeV is preferred with $g_\mu=3.8\times 10^{-4}$.
Following consideration of the constraints, the viable parameter space is severely limited, particularly by supernova constraints and nuclear physics, as will be seen in the following. Future experiments, such as NA62 and direct detection, will be able to test our model.

There are numerous constraints imposed by nuclear physics and experiments across a wide range.
\begin{itemize}

\item Supernova (SN 1987A) constraints for muon coupling $g_{\mu}$: cooling and energy explosion

In Ref.~\cite{Caputo:2021rux}, The authors were able to obtain all of the reliable constraints for scalars in a supernova. The general constraints are of two types: those resulting from the energy loss process and those resulting from the explosion energy. We finally determine the Yukawa couplings between mediator and muon based on the energy loss calculation, 
\begin{equation}
g_{\mu}>0.84 \times 10^{-4}.
\end{equation}
While muon $g-2$ discrepancy requires
\begin{equation}
g_{\mu} \simeq 0.4 \times 10^{-3}.
\label{eqn:g-2}
\end{equation}
Therefore muon $g-2$ anomaly is compatible with the energy loss process in a supernova. The loop-induced two-photon coupling, on the other hand, results in a diffuse  $\gamma$--ray constraint. We can avoid it by requiring the boson decay to occur within the radius of the progenitor star. This type of decay results in explosion energy within the progenitor star. The average explosion energy is some $10^{51}\mathrm{erg}$, so less than $1\%$ of the total energy release shows up in the explosion. To reduce the boson energy release to a level below $1\%$ of the explosion energy, we require
\begin{equation}
G_{\gamma \gamma}>5.3(4.8) \times 10^{-5} \mathrm{GeV}^{-1},
\end{equation}
where
\begin{equation}
G_{\gamma \gamma}=\frac{2 \alpha}{3 \pi} \frac{g_{\mu}}{m_{\mu}} B_{\phi}(x),
\end{equation}
and the loop factor $B$ is a function of $x=m_{\phi}/2m_{\mu}$,
\begin{equation}
B_{\phi}(x)=\frac{3}{2 x^{4}}\left[x^{2}-\left(1-x^{2}\right) \arcsin^{2}(x)\right].
\end{equation}
The corresponding Yukawa coupling is
\begin{equation}
g_{\mu}>3.6(3.3) \times 10^{-3}.
\end{equation}
This bound excludes the$(g-2)_{\mu}$ explanation for the muon coupling because it is an order of magnitude more restrictive than the required value in Eq.\ref{eqn:g-2}. Even we relax the bound by requiring $10\%$ energy deposition, the corresponding bound is $g_{\mu}>0.81\times 10^{-3}$, still excluding the $(g-2)_{\mu}$ explanation. The conclusion, however, is limited to the underlying assumption that the mediator only couples to muons. We can modify it by introducing the down-quark coupling $g_d$. In that case, the effective two-photon coupling becomes,
\begin{equation}
    G_{\gamma \gamma}=\frac{2 \alpha}{3 \pi} \frac{g_{\mu}}{m_{\mu}} B_{\phi}(m_{\phi}/2m_{\mu})+\frac{2Q_d^2\alpha}{3\pi}\frac{g_d}{m_d}B_{\phi}(m_{\phi}/2m_{d}),
\end{equation}
where $Q_d=1/3$. The upper bound of $g_d$ is determined by MiniBooNE, i.e.,$g_d\sim 10^{-3}$. In this case, we use a moderate value of $g_d$ to observe the shift of $g_{\mu}$  in the explosion energy constraint. For example, we take $g_d=2\times 10^{-4}$ as benchmark point. The $1\%$ energy deposition requires the corresponding $g_{\mu}$ to be
\begin{equation}
    g_{\mu}>2.7\times 10^{-3}.
\end{equation}
Meanwhile, the $10\%$ energy deposition results in a soft constraint,
\begin{equation}
    g_{\mu}>0.28\times 10^{-3}.
\end{equation}
As a result, the constraint from energy loss is not strict for muon $g-2$ anomaly, whereas the constraint from energy explosion is uncertain, excluding the possibility of muon $(g-2)$ anomaly in $1\%$  deposition while retaining hope in $10\%$ energy deposition. The additional issue we should care about is that, the existence of down-quark coupling not only shifts the two-photon coupling constraint for $g_{\mu}$ but re-introduces the energy loss process from nucleon-nucleon bremsstrahlung~\cite{Carenza:2019pxu}.
\begin{equation}
g_{a N}^{2} \equiv g_{a n}^{2}+0.61 g_{a p}^{2}+0.53 g_{a n} g_{a p} <8.26 \times 10^{-19}.
\end{equation}
Axions/scalars may be trapped in the SN core if the axion/scalar coupling to nucleons is strong enough. The cooling efficiency is reduced in this case, and the axion/scalar bound is weakened. The novel aspect of this constraint is that it only applies to the free-streaming regime. It has no bearing on the MeV scale force. Therefore we can safely ignore the supernova energy loss constraint on $g_d$.

    \item Nucleon Binding Energy in (N=Z) Nuclear Matter ($\delta B$ in NM)

The scalar coupling to nucleons contributes to the binding energy of nucleons in (N=Z) nuclear matter. In infinite nuclear matter, the change in nucleon binding energy is  \cite{Liu:2016qwd}
\begin{equation}
\delta B= \frac{(g_p+g_n)^2\rho}{4m_\phi^2},
\end{equation}
where $\rho\approx0.08{\rm\;fm^{-3}}$ is the nucleon density.
To avoid conflicts with nuclear physics, 
the extra contribution is less than 1 MeV.
It turns out that the constraints on the mediator mass $m_\phi$ and coupling $g_d$ are very strict. The mass range is limited to $m_\phi >4\times 10^{-1}$ MeV, and also vast space of $g_d$ is excluded as shown in black dashed line in Fig.~\ref{fig:gd}.

    \item $^3$He-$^3$H binding energy difference ($^3$He-$^3$H)

The binding energy difference between $^3$He and $^3$H, 763.76~keV was calculated using high precision by combining the contributions from Coulomb interaction and charge asymmetry of nuclear forces.  ~\cite{Friar:1969zz, Friar:1978mr, Coon:1987kt, Miller:1990iz, Wiringa:2013fia}.
So non-SM effects from the scalar mediator $\phi$ should be constrained, and it is set to be less than 30~keV~\cite{Liu:2016qwd}.
The upper limits on $g_d$ in our model are shown by the green dashed line in Fig.~\ref{fig:gd}.

    \item Neutron-Nucleus Scattering (N-N)

The low energy neutron-nucleus scattering experiment with the nucleus $^{208}$Pb constrains the coupling between the scalar mediator and neutron $g_N$~\cite{Leeb:1992qf}.
The constraints on coupling $g_d$ in our model are the replacement
$g_N^2/4\pi \rightarrow  1/4\pi (1-Z/A) g_n^2+(Z/A) g_p g_n$,
The scattered nucleus has an atomic mass of $A$ and an atomic number of $Z$.
In figure  \ref{fig:gd} , the purple dashed line represents the upper limits on $g_d$ as a function of the mediator mass $m_\phi$.
   \item MiniBooNE-Proton Beam Dumps

Fermilab Booster Neutrino Beam with an energy of 8 GeV probes the dark sector. Parameter space of dark photon is excluded between $5$ MeV and $50$ MeV~\cite{Aguilar-Arevalo:2018wea}. The exclusion for coupling $g_B$ maps into the coupling for $g_d$, where $g_d$ larger than $10^{-3}$ is ruled out. The boundary of the exclusion region is labeled by an orange dash-dotted line.

 \item Belle II constraints on $\mu\mu$+MET

Because of the muonic coupling, the mediator$\phi$ can be produced at $e^+e^-$  colliders in the process $e^+e^-\to\mu^+\mu^-\phi$, which then becomes $\mu^+\mu^- +\mathrm{MET}$final states. Recently, Belle II experiment performed the searches for the invisible decay of a $Z^\prime$ particle in the process $e^+e^-\to\mu^+\mu^-Z^\prime\to \mu^+\mu^- +{\rm MET}$ using 276 pb$^{-1}$ data \cite{Adachi:2019otg}, the null result put an upper bound on the muonic coupling which can be translated to be about $g_\mu<5\times 10^{-2}$ at the region of $m_{\phi} < 10^2$ MeV. And this constraint does not apply to our parameter region of interest. 
\item Beam-dump Constraints

Many neutral or charged mesons that can decay into photons and muons are produced in a fixed-target environment. In our model, with appropriate masses and interactions, they can decay into dark matter instead.
The constraint in E137~\cite{PhysRevD.38.3375} at SLAC is tiny in our model due to the existence of down quark coupling.
The invisible decay constraint $K^{+}\rightarrow \pi^+ \nu\nu$ in E949~\cite{Artamonov:2009sz} is negligible because our mediator only couples to down quarks. The appearance of up-type quark coupling at one-loop can reintroduce the beam-dump experiment constraint. However, because of its small couplings, it is less relevant in the recent experiment.
Finally, the invisible decaying scenario can help us probe our model in NA62 via leptonic decay $K\rightarrow \mu\nu\phi$. 
The excluded region will be determined by the following limits with existing data, ~\cite{CortinaGil:2021gga}
  \begin{equation}
      \mathrm{BR}(K^{+}\rightarrow \mu^+3\nu)<10^{-6}.
  \end{equation}
However, this will not affect our parameter space.
We also consider the future NA62 projected limits \cite{Krnjaic:2019rsv}, which is labeled by a magenta dashed line.

  \item Xenon1T constraints

  Accelerated dark matter is one scenario for detecting our sub-GeV dark matter. The most common example is cosmic-ray boosted dark matter caused by $\phi$-proton coupling. The differential event rate is explicitly stated
\begin{equation}
    \frac{dR}{dE_R}=\int_{T_{\chi}\left(T_{\chi}^{z, \min }\right)}^{\infty} d T_{\chi} \frac{d \sigma_{\chi N}}{d E_{R}} \frac{d \Phi_{\chi}}{d T_{\chi}},
    \label{eqn:event}
\end{equation}
where $T_{\chi}^{\min}$ is the minimal incoming dark matter kinetic energy to generate given recoil energy $E_R$. $d\sigma_{\chi N}/dE_R$ is differential cross-section between dark matter and nucleons, and $d\Phi_{\chi}/dT_{\chi}$ is differential dark matter flux. We can recover the sensitivity of a noble liquid detector on the sub-GeV dark matter using ~\ref{eqn:event}. Collisions with the local interstellar medium produce the local population of cosmic-ray dark matter. The attenuated dark flux can provide a few different ways to constrain the dark matter-nucleon cross-section in terms of ~\ref{eqn:event}. Implementing the momentum transfer effect can increase the event rate. We plot the exclusion limit in Fig.~\ref{fig:gd} with $m_{\phi}=0.1 m_{\chi}$.
The XenonNT and PandaX upgrades may push the exclusion line towards our model.
\end{itemize}

\section{Dark matter production}
\label{sec:DM}

Because dark matter is heavier than $3.4$ MeV mediator from the requirement of  $\Delta N_{\mathrm{eff}}$ and BBN \cite{Yeh:2020mgl}, direct annihilation into muons or pions is suppressed in comparison to the secluded scenario $\chi\chi\rightarrow \phi\phi$.
\begin{equation}
\sigma v_{\mathrm{rel}}=\frac{3 g_{\chi}^{4} v_{\text {rel }}^{2}}{128 \pi m_{\chi}^{2}},
\end{equation}
where $g_{\chi}$ is the coupling between dark matter and mediator in equation~\ref{eqn:Dark}. It is a P-wave process that is not constrained by CMB. When the Bohr radius of dark matter $1/\alpha_{\chi}m_{\chi}$ is less than the force radius $1/m_{\phi}$, Sommerfeld correction is unavoidable:
\begin{equation}
\sigma v_{\mathrm{rel}}=S_1\left(v_{\mathrm{rel}}\right) \times\left(\sigma v_{\mathrm{rel}}\right)_{0}.
\end{equation}

Under Hulthen approximation, the Sommerfeld enhancement is

\begin{equation}
S_{1}^{H}=\frac{\left(1-\varepsilon_{\phi} \pi^{2} / 6\right)^{2}+4 \varepsilon_{v}^{2}}{\left(\varepsilon_{\phi} \pi^{2} / 6\right)^{2}+4 \varepsilon_{v}^{2}} S_{0}^{H},
\end{equation}
where $\varepsilon_{v}=v_{\mathrm{rel}} / 2\alpha_{\chi}, \varepsilon_{\phi}=m_{\phi} /\left(\alpha_{\chi} m_{\chi}\right)$, $\alpha_{\chi}=g_{\chi}^2/4\pi$, and $S_0^{H}$ is s-wave Sommerfeld enhancement factor,

\begin{equation}
S_{0}^{H}=\frac{\pi}{\varepsilon_{v}} \frac{\sinh \left(\frac{2 \pi \varepsilon_{v}}{\pi^{2} \varepsilon_{\phi} / 6}\right)}{\cosh \left(\frac{2 \pi \varepsilon_{v}}{\pi^{2} \varepsilon_{\phi} / 6}\right)-\cos \left(2 \pi \sqrt{\frac{1}{\pi^{2} \varepsilon_{\phi} / 6}-\frac{\varepsilon_{v}^{2}}{\left(\pi^{2} \varepsilon_{\phi} / 6\right)^{2}}}\right)}.
\end{equation}

The thermal freeze-out process is calculated by solving the Boltzmann equation, which is shown below, 
\begin{equation}
\dot{n}+3 H n=-\langle\sigma v\rangle\left[n^{2}-n_{\mathrm{eq}}^{2}\right],
\label{eqn:Boltzmann1}
\end{equation}
where $\langle\sigma v\rangle$ represents the thermal average of the Sommerfeld corrected cross-section and $n_{\mathrm{eq}}$ represents the dark matter number density in chemical equilibrium.It is easy to solve the Boltzmann equation~\ref{eqn:Boltzmann1} with yield $Y_{\chi}=n_{\chi}/s$ as variable,

\begin{equation}
\frac{\mathrm{d} Y_{\chi}}{\mathrm{~d} x}=-\frac{\langle\sigma v\rangle}{H_{\mathrm{eff}} x} s_{\mathrm{SM}}\left(Y_{\chi}^{2}-Y_{\mathrm{eq}}^{2}\right).
\label{eqn:Boltzmann2}
\end{equation}

In most cases, the P-wave process is CMB-safe. However, as discussed in \cite{An:2016kie} the bound state formation cross-section is S-wave, reintroducing the CMB constraint. The monopole transition into the S-wave bound state can be solved analytically, 
 \begin{equation}
(\sigma v)_{n 0}^{M}=\frac{2^{6} \pi^{3} \alpha_{\chi}^{5} e^{-4 n}\left(L_{n-1}^{1}(4 n)\right)^{2}}{9 n^{3} m_{\chi} m_{\phi} \sin ^{2}\left(\pi \sqrt{\alpha_{\chi} m_{\chi}} / m_{\phi}\right)}.
\label{eqn:BSF1}
\end{equation}
The contribution in equation~\ref{eqn:BSF1} is significant only for CMB, but not for relic density, because its leading-order contribution cancels out for the real scalar mediator. The solution of equation ~\ref{eqn:Boltzmann2} is referred to as the minimal benchmark model of dark matter, which is denoted as Case-1. The minimal setup variations are classified as follows:
\begin{enumerate}[(1)]
    \item The reason why bound state formation is so small in comparison to annihilation cross-section is that the real scalar mediator cancels out in the radiative formation of bound state. Given this, we propose a modification to Case-1, Case-2:
    \begin{equation}
        - \delta \mathcal{L}=c_{\phi} \phi^3
    \end{equation}

 which is the mediator's cubic self-interaction As a result, one-mediator emission of bound-state formation is non-vanishing \cite{Oncala:2019yvj,Binder:2019erp,Oncala:2018bvl},

\begin{align}
\sigma_{\mathrm{BSF}} v_{\mathrm{rel}}&= \frac{\alpha_{\chi}^{2}}{\mu^{2}}\left(\frac{2 c_{\phi}}{\mu \alpha_{\chi}^{2}}\right)^{2} \sqrt{1-\left(\frac{2 m_{\phi}}{\left(\alpha_{\chi}^{2}+v_{\mathrm{rel}^{2}}\right)}\right)^{2}} S_{0}^{H}(\varepsilon_v, \varepsilon_{\phi})\nonumber \\
&\times\left(\frac{1/\varepsilon_{\phi}^{2}}{1+1/\varepsilon_{\phi}^{2}}\right)^{2} \exp (-4/\varepsilon_{\phi} \operatorname{arccot}(1/\varepsilon_{\phi})).
\end{align}

In this case, the Boltzmann equation becomes coupled, and the bound state's number density evolution is observed. When we generalize the definition of the effective cross-section, it can be reduced to a single Boltzmann equation.

\begin{equation}
\left\langle\sigma_{\text {eff }} v_{\text {rel }}\right\rangle=\left\langle\sigma_{\text {ann }} v_{\text {rel }}\right\rangle+\left\langle\sigma_{\text {BSF }} v_{\text {rel }}\right\rangle_{\text {eff }},
\end{equation}
where the effective bound state formation scales as a function of decay, bound state ionization,
\begin{equation}
\left\langle\sigma_{\mathrm{BSF}} v_{\mathrm{rel}}\right\rangle_{\mathrm{eff}}=\left\langle\sigma_{\mathrm{BSF}} v_{\mathrm{rel}}\right\rangle \times\left(\frac{\Gamma_{\mathrm{dec}}}{\Gamma_{\mathrm{dec}}+\Gamma_{\mathrm{ion}}}\right),
\end{equation}
with the decay and ionization of the bound state being
\begin{equation}
\Gamma_{\mathrm{decay}}=\frac{0.01\left|\Psi_{100}(0)\right|^{2} g_{\chi}^8}{49152 \pi^{6} m_{\chi}^{2}},\quad
\Gamma_{\mathrm{ion}} =\left\langle\sigma_{\mathrm{BSF}} v_{\mathrm{rel}}\right\rangle\left(\frac{m_{\chi} T}{4 \pi}\right)^{3 / 2} e^{-\left|E_{\mathrm{B}}\right| / T}.
\end{equation}

It is easy to find that the cubic term $c_{\phi}$ must be larger than $10^{4}\mathrm{GeV}$ in our parameter space to affect relic density. However, such a large cubic coupling complicates CMB ionization and vacuum stability. As a result, Case-2 is unimportant in our model. The detailed analysis of Case-2 will be left for a future study.

\item The derivation of equation~\ref{eqn:Boltzmann1} has an underlying assumption: During chemical freeze-out, kinetic equilibrium persists. It becomes constrained at a certain parameter space where elastic scattering between dark matter and mediator is insufficient to maintain kinetic equilibrium. It implies that additional variables must be included to complete the equation. The convenient approach is to consider dark matter temperature $T_{\chi} \equiv g_{\chi} /\left(3 n_{\chi}\right) \int d^{3} p(2 \pi)^{-3}\left(p^{2} / E\right) f_{\chi}$.
For the two variables system $n_{\chi}$, $T_{\chi}$, the coupled Boltzmann equation becomes

\begin{equation}
\begin{aligned}
\frac{Y^{\prime}}{Y}=& \frac{s Y}{x \tilde{H}}\left[\frac{Y_{\mathrm{eq}}^{2}}{Y^{2}}\langle\sigma v\rangle_{T}-\langle\sigma v\rangle_{T_{\chi}}\right],  \\
\frac{y^{\prime}}{y}=& \frac{1}{x \tilde{H}}\left\langle C_{\mathrm{el}}\right\rangle_{2}+\frac{s Y}{x \tilde{H}}\left[\langle\sigma v\rangle_{T_{\chi}}-\langle\sigma v\rangle_{2, T_{\chi}}\right] \\
&+\frac{s Y}{x \tilde{H}} \frac{Y_{\mathrm{eq}}^{2}}{Y^{2}}\left[\frac{y_{\mathrm{eq}}}{y}\langle\sigma v\rangle_{2, T}-\langle\sigma v\rangle_{T}\right]+2(1-w) \frac{H}{x \tilde{H}},
\end{aligned}
\end{equation}
where $Y(x) \equiv n / s, y(x) \equiv m_{\chi} T_{\chi} s^{-2 / 3}$. We label this modification as Case-3.

When calculating kinetic decoupling, the temperature of dark matter is not always the same as the temperature of the thermal bath, which becomes significant when the elastic scattering between dark matter and mediator is inefficient. The quantity to capture the elastic scattering process is $C_{\mathrm{el}}$ in the Fokker-Plank operator

\begin{equation}
C_{\mathrm{el}}=\frac{E}{2} \gamma(T)\left[T E \partial_{p}^{2}+\left(2 T \frac{E}{p}+p+T \frac{p}{E}\right) \partial_{p}+3\right] f_{\chi},
\end{equation}
where $\gamma$ is the momentum transfer rate,

\begin{equation}
\gamma=\frac{1}{3 g_{\chi} m_{\chi} T} \int \frac{\mathrm{d}^{3} k}{(2 \pi)^{3}} g^{\pm}(\omega)\left[1 \mp g^{\pm}(\omega)\right] \int_{-4 k_{\mathrm{cm}}^{2}}^{0} \mathrm{~d} t(-t) \frac{\mathrm{d} \sigma}{\mathrm{d} t} v.
\end{equation}

In our model, the differential cross-section corresponds to $\chi\phi\rightarrow\chi\phi$ process with $(\mathrm{d} \sigma / \mathrm{d} t) v \equiv|\mathcal{M}|_{\chi f \leftrightarrow \chi f}^{2} /\left(64 \pi k \omega m_{\chi}^{2}\right)$, which can be given explicitly,

\begin{align}
  &\int\mathrm{d}t (-t) \mathrm{d} \sigma / \mathrm{d} t) v
  =\frac{1}{3(m_{\phi}^2+2\omega m_{\chi})^2} \times\nonumber\\
  &\frac{\sum_i (4k^2m_{\chi}^2)^{i-1}C_i}{\left.\left(m_{\phi}^{4}+\left(4(k-\omega)(k+\omega)+m_{\phi}^{2}\right) m_{\chi}^{2}-2 \omega m_{\chi}^{3}\right)\left(m_{\phi}^{2}+m_{\chi}\left(2 \omega+m_{\chi}\right)\right)^{3}\right)},
\end{align}
where $C_i$ is just combination of $m_{\chi}, m_{\phi}$ and $\omega$:
\begin{align}
C_1&=-3 \left(\log \left[-m_{\phi}^{2}+2 \omega m_{\chi}\right]-\log \left[-m_{\phi}^{2}+2 m_{\chi}\left(\omega-\frac{2 k^{2} m_{\chi}}{m_{\phi}^{2}+2 \omega m_{\chi}+m_{\chi}^{2}}\right)\right]\right)\nonumber\\
& \times\left(m_{\phi}^{2}+2 \omega m_{\chi}\right)
(m_{\phi}^{6}-2 m_{\phi}^{4} m_{\chi}^{2}-m_{\phi}^{2} m_{\chi}^{2}\left(\omega^{2}+4 m_{\chi}\left(3 \omega+m_{\chi}\right)\right)\nonumber\\
&-2 \omega m_{\chi}^{3}\left(\omega^{2}-4 m_{\chi}\left(2 \omega+3 m_{\chi}\right)\right)),
\end{align}

\begin{align}
C_2&=-16k^6m_{\chi}^6(m_{\phi}^2+2\omega m_{\chi})+6k^2 m_{\chi}^2(m_{\phi}^2+m_{\chi}(2\omega+m_{\chi}))^2\nonumber\\
&\times (-m_{\phi}^6+2m_{\phi}^4 m_{\chi}(m_{\chi}-2\omega)+2\omega m_{\chi}^3(-5\omega^2 +8\omega m_{\chi}+12m_{\chi}^2)\nonumber\\
&+m_{\phi}^2m_{\chi}^2(-7\omega^2+20m_{\chi}(\omega+m_{\chi}))) + 8k^4 m_{\chi}^4(m_{\phi}^2+m_{\chi}(2\omega+m_{\chi}))\nonumber\\
&\times (m_{\phi}^4+6m_{\phi}^2 m_{\chi}(\omega+m_{\chi})+m_{\chi}^2(11\omega^2+12m_{\chi}(2\omega+m_{\chi})))\nonumber\\
&-3(m_{\phi}^2+2\omega m_{\chi})(m_{\phi}^2+m_{\chi}(2\omega+m_{\chi}))^3  (m_{\phi}^6-2m_{\phi}^4m_{\chi}^2-m_{\phi}^2m_{\chi}^2\nonumber\\
&\times (\omega^2+4m_{\chi}(3\omega+m_{\chi}))-2\omega m_{\chi}^3(\omega^2-4m_{\chi}(2\omega+3m_{\chi}))).
\end{align}

\end{enumerate}

\begin{figure}[!htbp]
\centering
\includegraphics[width=0.46\textwidth]{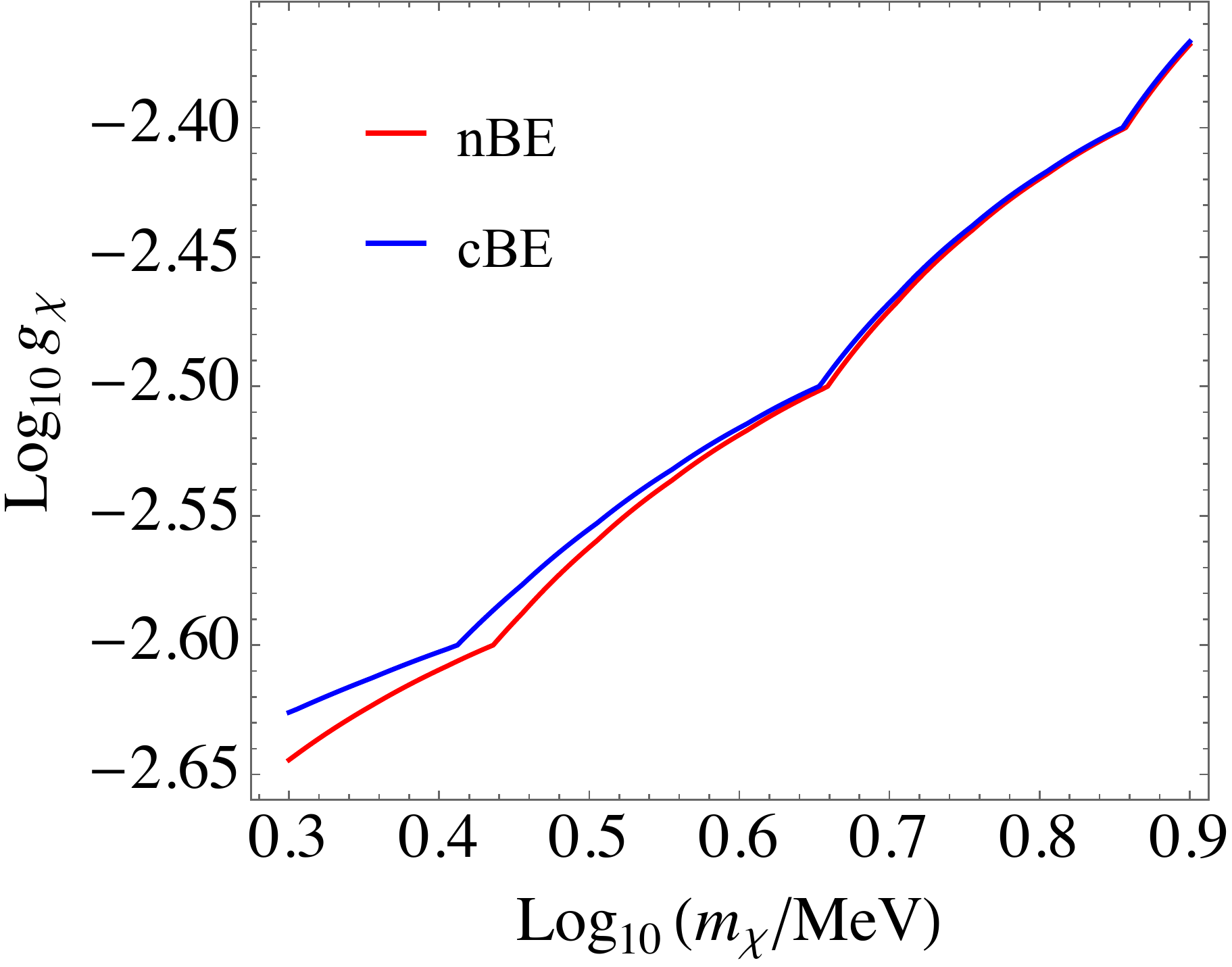}
\includegraphics[width=0.51\textwidth]{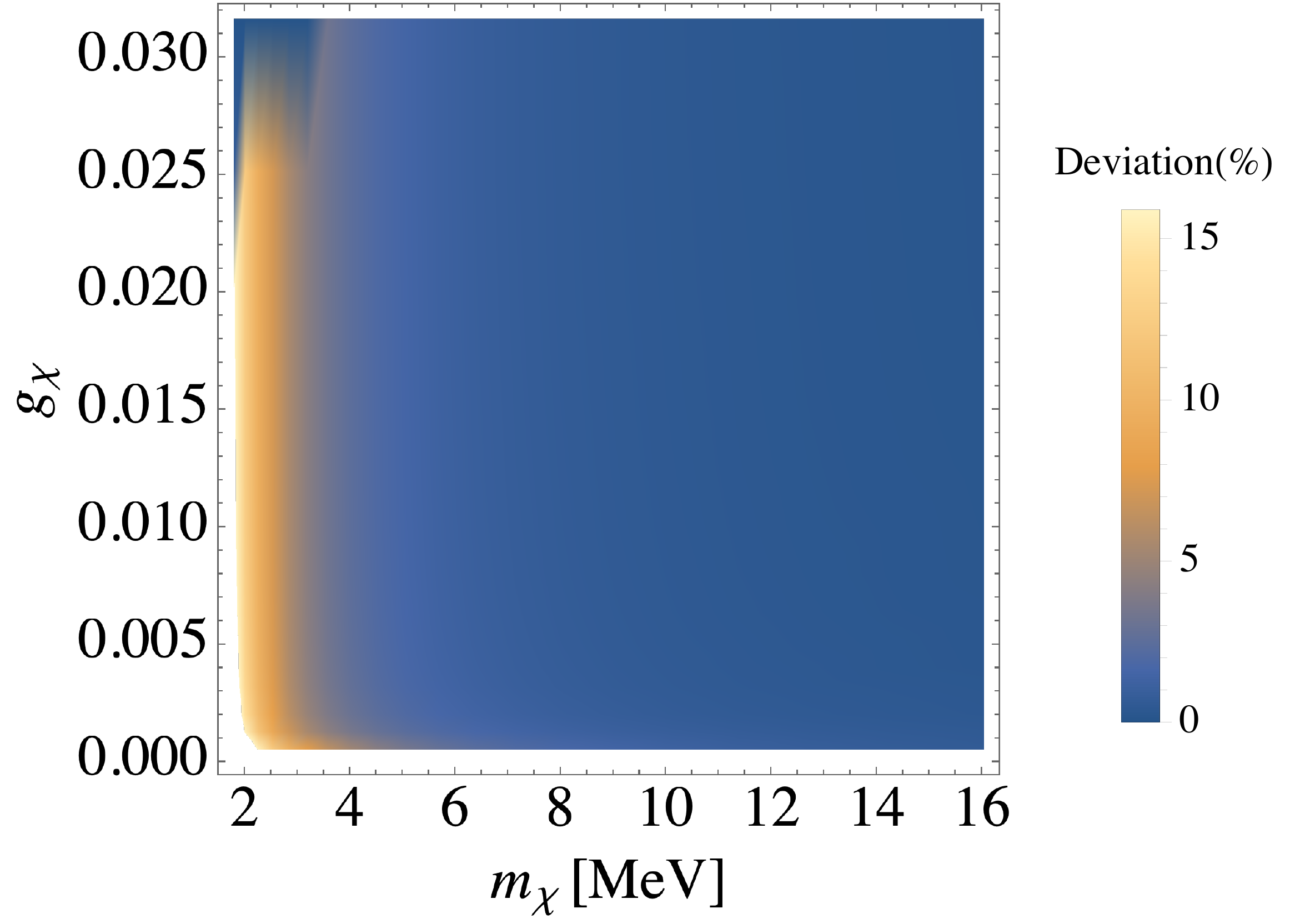}
\caption{Left: Correct DM relic density for Case-1 (nBE) and Case-3 (cBE) in ($m_{\chi}-g_{\chi}$) plane. We fixed $m_\phi=3.4$ MeV.
Right: the comparison of Case-1 and Case-3, the deviation is defined as  $(\Omega_{\chi}^{\rm cBE}h^2-\Omega_{\chi}^{\rm nBE}h^2)/\Omega_{\chi}^{\rm cBE}h^2$.}
\label{fig:relic}
\end{figure}

In Fig.~\ref{fig:relic}, we show the correct relic density in $m_{\chi}-g_{\chi}$ plane.  We use DRAKE~\cite{Binder:2021bmg} to compute the relic density in Case-1 (nBE) and Case-3 (cBE) in the left panel of Fig.~\ref{fig:relic}. The red line represents Case-1's correct relic density, while the blue line represents Case-3 with the early kinetic decoupling effect. We can see that the deviation occurs when a small mass splits between dark matter and the mediator. When the dark matter becomes significantly heavier than the mediator, the two scenarios degenerate. It can be seen in right panel the deviation (defined as $(\Omega_{\chi}^{\rm cBE}h^2-\Omega_{\chi}^{\rm nBE}h^2)/\Omega_{\chi}^{\rm cBE}h^2$) can be at most $15\%$.

\section{Conclusion}
We propose a flavor-specific scalar mediator model with only muon and down-quark coupling in this paper. These appear to be viable explanations for the muon g-2 anomaly and the proton radius puzzle. The interaction with muons and down quarks results in observations in various experiments, which can constrain our model. The MeV mediator has loop-induced coupling with photons. Therefore it leads to observational phenomenologies in a supernova. In particular, the $1\%$ energy deposition into progenitor star excludes the possibility of the muon g-2 discrepancy. However,  $10\%$ energy deposition remains the best hope for resolving the disparity. We also discover that there is little parameter space left for $g_{\mu}$ and $g_{d}$ implying that precision measurements on proton radius and other high-intensity frontiers such as NA62 have significant future discovery potential for our model. To produce the right amount of thermal relics during cosmological evolution,  we study all the possible modifications to the secluded dark matter. For MeV-scale dark matter in the early kinetic decoupling modification, the deviation in relic density can be as high as $10\%$. While considering the vacuum stability requirements and the CMB constraint, the bound state formation modification is negligible.

\section*{Acknowledgements}

Bin Zhu was supported by the National Natural Science Foundation of China under the grants No. 11805161, by the Natural Science Foundation of Shandong Province under the grants No. ZR2018QA007, by the Basic Science Research Program through the National Research Foundation of Korea (NRF) funded by the Ministry of Education, Science and Technology (NRF-2019R1A2C2003738), and by the Korea Research Fellowship Program through the NRF funded by the Ministry of Science and ICT (2019H1D3A1A01070937).  
Xuewen Liu was supported by the National Natural Science Foundation of China under the Grants No. 11947034 and No. 12005180, and by the Natural Science Foundation of Shandong Province under the Grant No. ZR2020QA083.
This work is  also supported by the Project of Shandong Province
Higher Educational Science and Technology Program under Grants No. 2019KJJ007.

\bibliographystyle{JHEP.bst}
\bibliography{lit}

\end{document}